\newif\ifsubmission   %
\newif\iffinal        %

\submissionfalse 
\finalfalse

\newif\iffull
\ifsubmission \fullfalse
\else \iffinal \fullfalse
\else \fulltrue
\fi\fi

\ifsubmission
\documentclass[runningheads,11pt]{llncs}
\usepackage[letterpaper,hmargin=1in,vmargin=1.25in]{geometry}
\else
\iffinal
\documentclass[runningheads]{llncs}
\else
\documentclass[runningheads,11pt]{llncs}
\usepackage[letterpaper,hmargin=1in,vmargin=1.25in]{geometry}
\fi
\fi

\usepackage{graphicx,amsmath,amssymb,xspace}  
\usepackage{xspace} 
\usepackage{wrapfig}
\usepackage{citesort}

\newcommand{\ourtitle}{
An Enciphering Scheme Based on a Card Shuffle
}

\iffinal
\authorrunning{Viet Tung Hoang, Ben Morris, and Phillip Rogaway}
\fi
\titlerunning{\ourtitle} 


\DeclareMathAlphabet{\mathsl}{OT1}{cmr}{m}{sl}
\DeclareMathAlphabet{\mathsc}{OT1}{cmr}{m}{sc}

\newcommand{\name}{\mbox{swap-or-not}\xspace}  %
\newcommand{\Name}{\mbox{Swap-or-not}\xspace}  %
\newcommand{\NAME}{\mbox{Swap-Or-Not}\xspace}  

\newcommand{\blockcipher}{blockcipher\xspace}
\newcommand{\blockciphers}{blockciphers\xspace}

\newcommand{\Blockciphers}{Blockciphers\xspace}
\newcommand{\comment}[1]{~\mbox{\small/\hspace{-0.7ex}/\textrm{#1}}}
\newcommand{\commentt}[1]{~~~\mbox{\small/\hspace{-0.7ex}/\textrm{#1}}}
\newcommand{\eg}{e.g.}
\newcommand{\ie}{i.e.}

\newcommand{\SN}{\mathrm{SN}}   
\newcommand{\Sh}{\mathrm{SN}}   
\newcommand{\TSN}{\mathrm{TSN}}
\newcommand{\TSpace}{{\mathcal T}}
\newcommand{\tweak}[1]{\widetilde{#1}}

\newcommand{\LL}{m}

\def\sfrac#1#2{{\textstyle{#1 \over #2}}}
\def \cards {{\cal C}}
\newcommand{\mT}{X}    %
\newcommand{\mtau}{\tau} %
\newcommand{\mpi}{\pi} %
\newcommand{\card}{{z}}

\newcommand{\n}{t}
\newcommand{\given}{\,| \;}

\newcommand{\TildeTheta}{\Tilde{\Theta}}

\def\half{{\textstyle{1\over2}}}
\def\quarter{{\textstyle{1\over4}}}

\newcommand{\deck}{W}   %

\newcommand{\Colon}{\!:\,}

\newcommand{\hatX}{\smash{\hat{X}}}
\newcommand{\ip}{\odot}  %

\newcommand{\nr}{r}

\newcommand{\F}{F}
\newcommand{\Exp}{\mathbf{E}} %

\newcommand{\cca}{\mathrm{cca}}

\newcommand{\ncpa}{\mathrm{ncpa}}

\newcommand{\Adv}{\mathbf{Adv}}
\newcommand{\NA}[1]{\Adv^\ncpa_{#1}}
\newcommand{\AD}[1]{\Adv^\cca_{#1}}

\newcommand{\NN}{[N]}

\newcommand{\xor}{\oplus}
\newcommand{\bits}{\{0,1\}}
\newcommand{\getsr}{{\:\stackrel{\scriptscriptstyle \hspace{0.2em}\$}{\leftarrow}\:}}

\newcommand{\calK}{{\cal K}}
\newcommand{\calM}{{\cal M}}

\newcommand{\calX}{{\cal X}}
\newcommand{\Perm}{\mathrm{Perm}}

\newcommand{\KF}{{K\!F}}
\newcommand{\KL}{{K\!L}}
\newcommand{\K}{K}
\newcommand{\X}{X}
\newcommand{\Y}{Y}
\newcommand{\Kappa}{K}

\newcommand{\Group}{G}
\newcommand{\code}[1]{\langle{#1}\rangle}

\newcommand{\Shuffle}{\Sh}

\newcommand{\captionfont}{\small}
\newcommand{\heading}[1]{\vspace{12pt}\noindent\textsc{#1} }
\newcommand{\noskipheading}[1]{\textsc{#1} }
\newcommand{\headingg}[1]{\noindent\textsc{#1} }

\newcommand{\figref}[1]{Fig.~\ref{#1}}
\newcommand{\lemref}[1]{Lemma~\ref{#1}}
\newcommand{\secref}[1]{Section~\ref{#1}}
\newcommand{\thref}[1]{Theorem~\ref{#1}}
\newtheorem{lemmaa}[theorem]{Lemma}

\newlength{\saveparindent}
\newlength{\saveparskip}
\iffinal
\title{\ourtitle}
\else
\title{\Large\bf \ourtitle} 
\fi

\ifsubmission
\author{\ } 
\institute{\ } 
\else
\author{Viet Tung Hoang\inst{1} \quad Ben Morris\inst{2} \quad Phillip Rogaway\inst{1}}           
\institute{Dept.\ of Computer Science, University of California, Davis, USA \and
Dept.\ of Mathematics, University of California, Davis, USA}

\fi

\begin{document}

\maketitle
\iffull \centerline{\today} \fi
\vspace{-3ex}

\ifsubmission\vspace{-0.7in}\fi
\begin{abstract}
We introduce the \textit{\name shuffle} and 
show that the technique gives rise to a new method 
to convert a pseudorandom function (PRF)
into a pseudorandom permutation (PRP)
(or, alternatively, to directly build a confusion/diffusion \blockcipher). 
We then prove that \name has excellent quantitative security bounds, 
giving a Luby-Rackoff type result
that ensures security
(assuming an ideal round function)
to a number of adversarial queries 
that is nearly the size of the construction's domain.
\Name provides a direct solution for 
building a small-domain cipher
and achieving format-preserving encryption, 
yielding the best 
bounds known for a practical scheme for enciphering credit-card numbers.
The analysis of \name is based on the theory of mixing times of Markov chains.

                   \mbox{PRF-to-PRP} \mbox{conversion},  
                   pseudorandom permutations, \name. 

\end{abstract}

\section{Introduction} \label{se-intro}

\noskipheading{Overview.}
Despite the diversity of proposed \blockciphers, 
only two approaches underlie the construction of real-world designs:
essentially everything looks like some sort of 
{\mbox{Feistel} \mbox{network}} (\eg, DES, FEAL, MARS, RC6)
or 
\mbox{SP-network}  (\eg, Rijndael, Safer, Serpent, Square).
Analogously, in the literature on constructing 
pseudorandom permutations (PRPs) from pseudorandom functions (PRFs), 
we have provable-security analyses for Feistel variants (\eg, \cite{lr,nr,patarin,hr,hkt}),
as well as modes of operation (\eg, \cite{nr,nr2,cmc,eme2}) that can
again be construed as \mbox{SP-networks}, now on a large domain.
Perhaps there just are not that many fundamentally different ways to make a \blockcipher.
Or perhaps we might have failed to notice \textit{other} possibilities. 

In this short paper we describe a very different way to 
make a \blockcipher. We call it a \textit{\name} network
(or {cipher} or {shuffle}).
Besides introducing the construction, we evidence its 
cryptographic utility. We do this by showing that \name 
provides the quantitatively best mechanism known, in terms of concrete security bounds,
to convert a PRF into a PRP.
We also show that \name provides a practical solution for 
the problem of format-preserving encryption (FPE) on domains of troublesome size,
such as enciphering credit-card numbers.

%
\begin{wrapfigure}{l}{3.4in}{%
\iffull \vspace{-4.8ex} \fi
\fbox{~\parbox{3.25in}{ %
\vspace{-2.7ex}  %
\begin{tabbing}
123\=123\=\kill  
proc $E_{\KF}(\X)$  \`\commentt{\bf \name$\!\!$}\\[0.5ex]
{for} $i\gets1$ {to} $\nr$ {do}\\
\iffull
\>   $\X'\gets \K_i\xor \X$, ~$\hatX \gets \max(\X,\X')$\\
\else
\>   $\X'\gets \K_i\xor \X$\\
\>   $\hatX \gets \max(\X,\X')$\\
\fi
\>   {if} $\F_i(\hatX)=1$ then $\X \gets \X'\;$\\ 
{return} $\X$
\end{tabbing}
\vspace{-2.2ex}
{\small \textbf{Fig.~1.}~~Cipher $E=\SN[r,n]$ encrypts \mbox{$\X\!\in\!\bits^n$} using 
a key $\KF$ naming $\K_1,\ldots, \K_\nr\!\in\!\bits^n$ and
round functions \mbox{$\F_1,\ldots,\F_\nr\Colon\bits^n\rightarrow\bits$}.}
}}}
\vspace{-4ex}
\end{wrapfigure}
\iffull \vspace{2ex} \fi
\noindent\textsc{Construction.}
Suppose we aim to encipher \mbox{$n$-bit} strings; our message space 
is the set \mbox{$\calX=\bits^n$}. 
Assume we will use~$\nr$ rounds, and that the \blockcipher's key~$\KF$ 
names subkeys \mbox{$\K_1,\ldots, \K_\nr\in\bits^n$} as well as 
round functions \mbox{$\F_1,\ldots,\F_r$}, each of which
maps $n$-bits to a single bit, so $F_i\Colon\bits^n\rightarrow\bits$.
Then we encipher \mbox{$\X\in\bits^n$} as shown in Fig.~1.
The reason that this works, that one gets a permutation, 
is simply that $\X\mapsto \K_i\xor \X$ is an involution, and our round function depends on the {set} $\{\X,\K_i\xor \X\}$.
The inverse direction for \name is identical to the forward direction shown above except 
for having~$i$ run from~$\nr$ down to~$1$.

Restating the algorithm in English, at each round~$i$ we pair the current value of~$\X\in{\bits^n}$ with a
``partner'' point $\X' = \K_i \xor \X$.  We either replace~$\X$ by its partner or leave it alone.
Which of these two things we do is determined by applying the boolean-valued~$\F_i$ to the two-element set $\{\X,\X'\}$. 
Actually, in order to give~$\F_i$ a more conventional domain, we select a canonical representative from 
$\{\X,\X'\}$, say $\hatX= \max(\X,\X')$, and apply~$\F_i$ to it.
Note that each plaintext
maps to a ciphertext by xoring into it some subset of
the subkeys $\{\K_1,\ldots, \K_\nr\}$.
This might sound linear, but it most definitely is not. 

\heading{Card shuffling view.}
The \name construction was invented, and will be analyzed, by regarding it as
a way to shuffle a deck of cards. 
Seeing a \blockcipher as a card shuffle enables one to exploit a large body
of mathematical techniques, these dating back to the first half of the twentieth century.
In addition, some ways to shuffle cards give rise to 
enciphering schemes that cryptographers did not consider. 
\Name is such a case.

One can always see a card shuffle as an enciphering scheme, and vice~versa.
If you have some method to shuffle~$N$ cards, this determines a corresponding way
to encipher~$N$ points: place a card at each position $\X\in\NN$, where $\NN=\{0,1,\ldots,N\!-\!1\}$;
shuffle the deck; then look to see the position where the card initially at position~$\X$ ended up. 
Call that position the ciphertext~$\Y$ for~$\X$. 
The randomness used in the shuffle corresponds the cipher's key.

The first thing needed for a card shuffle to give rise to a computationally feasible
\blockcipher is that the shuffle be \textit{oblivious}, an idea %
suggested by Moni~Naor \cite[p.~62]{nr}, \cite[p.~17]{rudich}.
In an {oblivious} shuffle one can trace the trajectory of a card
without attending to lots of \textit{other} cards in the deck. 
Most conventional shuffles, such as the riffle shuffle, are not oblivious.
The Thorp shuffle \cite{thorp} is oblivious---and so is \name.
As a shuffle, here's how it looks.

\iffull \else\newpage\fi
\begin{wrapfigure}{r}{3.2in}{%
\iffull \vspace{-4.5ex} \fi
\fbox{~\parbox{3in}{ %
\vspace{-2.5ex}
\begin{tabbing}
123\=123\=123\=\kill
$\Kappa\getsr \bits^n$ \`\commentt{\bf \name as a shuffle}\\
 for each pair of positions $\{\X, \Kappa \xor \X\}$ %
                 \\
 \>  $b\getsr\bits$\\
 \>  if $b=1$ then swap the cards\\
 \>  \quad at positions $\X$ and $\Kappa \xor \X$
\end{tabbing}
{\small \textbf{Fig.~2.}~~Mixing a deck of $N\!=\!2^n$ cards, 
each at a position $\X\in\bits^n$.
The code shows one shuffle. 
For better mixing, the shuffle is repeated $r$ times.
}
}}}
\vspace{-4ex}
\end{wrapfigure}
Recasting \name as a way to shuffle cards,
suppose we have~$N$ cards, one at each position
\mbox{$\X\in\NN$}, where \mbox{$N=2^n$}.
To shuffle the deck,
choose a random \mbox{$\Kappa\!\in\!\bits^n$} and then, for each pair of card positions~$\X$ and $\Kappa \xor \X$,
flip a fair coin. If it lands heads, swap the cards at the indicated positions;
if it lands tails, leave them alone. 
See Fig.~2.
The process can be repeated any number~$\nr$ times, 
using independent coins (both the $K$-values and the $b$-values) 
for each shuffle. 

When the \name shuffle of Fig.~2 is translated back into the language of encryption, 
one recovers the \name cipher of Fig.~1; these are different views of precisely the same process. 
The random pairing-up of cards specified by~$\Kappa$ for the $i$th shuffle corresponds 
to the subkey~$\K_i$. The random bit~$b$ flipped at the shuffle's round~$i$ 
for the pair $\{\X,\Kappa\xor \X\}$ corresponds $\F_i(\hatX)$.

\iffull \newpage \fi
\vspace{10pt}
\begin{wrapfigure}{l}{3.4in}{%
\fbox{\parbox{3.3in}{%
\vspace{-2.0ex}
\begin{tabbing}
123\=123\=\kill
 proc $E_{\KF}(\X)$ \`\comment{\bf Generalized domain} \\
 {for} $i\gets1$ {to} $\nr$ {do}\\
 \iffull
  \>   $\X'\gets \K_i - \X$, ~$\hatX \gets \max(\X,X')$\\
 \else
 \>   $\X'\gets \K_i - \X$\\
 \>   $\hatX \gets \max(\X,X')$\\
 \fi
 \>   {if} $\F_i(\hatX)=1$ then $\X \gets \X'\;$\\ 
{return} $\X$ 
\end{tabbing}
{\small \textbf{Fig.~3.}~~Cipher $E=\SN[r,N,+]$ encrypts \mbox{$\X\!\in\!\NN$} using 
a key $\KF$ naming $\K_1,\ldots, \K_\nr\!\in\![N]$ and round functions 
\mbox{$\F_1,\ldots,\F_\nr\Colon[N]\rightarrow\bits$}.}
}}}
\vspace{-4ex}
\end{wrapfigure}
\noindent\textsc{Generalizing.}
It is useful to be a bit more general here,
working in a finite abelian 
group \mbox{$\Group=(\NN, +)$} instead of the group $(\bits^n,\xor)$ of bit strings under xor. 
(For convenience, we have assumed that the group elements are
named $[N]=\{0,\ldots,N-1\}$.)
In this way we won't need the number of points~$N$ in the message space 
$\calX=[N]$ to be a power of two---we'll be able to encipher points on 
any set~$\calX=\NN$, just by naming a group operator, say addition modulo~$N$.
For generalizing the shuffle of Fig.~2, the value~$\Kappa$ is uniformly drawn from~$\NN$
rather than from $\bits^n$, and
we consider the pair of positions $\{\X,\Kappa-\X\}$
rather than $\{\X,\Kappa\xor \X\}$.
For the generalized cipher---see Fig.~3---the  
key~$\KF$ will name subkeys $\K_1,\ldots, \K_\nr\in\NN$ and round functions 
$\F_1,\ldots, \F_\nr\Colon\NN\rightarrow\bits$. We set $\X'\gets \K_i-\X$ rather
than $\X'\gets \K_i\xor \X$. 
The inverse remains what one gets by iterating from~$r$ down to~$1$.

\heading{Results.}  
As with Luby and Rackoff's seminal paper~\cite{lr}, 
we can analyze the \name construction 
by regarding its constituent parts as uniformly random.  
Formally, let us write $\SN[r,N,+]\colon\calK\times[N]\rightarrow[N]$ 
for the \blockcipher~$E$ specified in Fig.~3 that is \name 
with~$r$ rounds, a message space of~$[N]$, the indicated group operator, and where
the key space names all possible \mbox{subkeys} \mbox{$\K_1,\ldots,\K_r\in[N]$} and 
all possible round functions $\F_1,\ldots,F_r\Colon[N]\rightarrow\bits$.
Thus a random key~$\KF$ for this cipher has the $\K_i$ and $\F_i$ values uniformly chosen.
We define the CCA (also called the ``strong-PRP'') 
advantage of an adversary~$A$ attacking~$E$ by 
dropping it into one of two worlds. 
In the first, the adversary gets an oracle for $E_\KF(\cdot)$, for a random~$\KF$,
and also an oracle for its inverse, $E_\KF^{-1}(\cdot)$.
Alternatively, the adversary is given a uniformly random permutation 
$\pi\Colon \NN\rightarrow\NN$, along with its inverse,~$\pi^{-1}(\cdot)$. 
Define 
\[\Adv^\cca_{\SN[r,N,+]}(q) \;\;=\;\; \max_A\; 
\left\{\Pr[A^{\,E_\KF(\cdot),\:E_\KF^{-1}(\cdot)}\Rightarrow1] - 
  \Pr[A^{\,\pi(\cdot),\:\pi^{-1}(\cdot)}\Rightarrow1]
\right\},
\]
the maximum over all adversaries that ask at most~$q$ total queries.
Our main result is that 
\begin{eqnarray}
\textbf{Adv}^{\mathrm{cca}}_{\SN[r,N,+]}(q) &\;\;\le\;\;& \frac{8N^{3/2}}{r + 4} \left(\frac{q + N}{2N}\right)^{r/4 + 1}\;.
\end{eqnarray}
Roughly said, you need $r=6\lg N$ rounds of \name to start to see a good bound on CCA-security.
After that, the adversary's advantage drops off inverse exponentially in~$r$. 
The summary explanation of formula~(1) 
just given assumes that the number of adversarial queries is 
capped at $q=(1-\epsilon)N$ for some fixed~$\epsilon>0$.  

\setcounter{figure}{3}
\begin{figure}[t]
\iffull
\scalebox{0.45}{\includegraphics{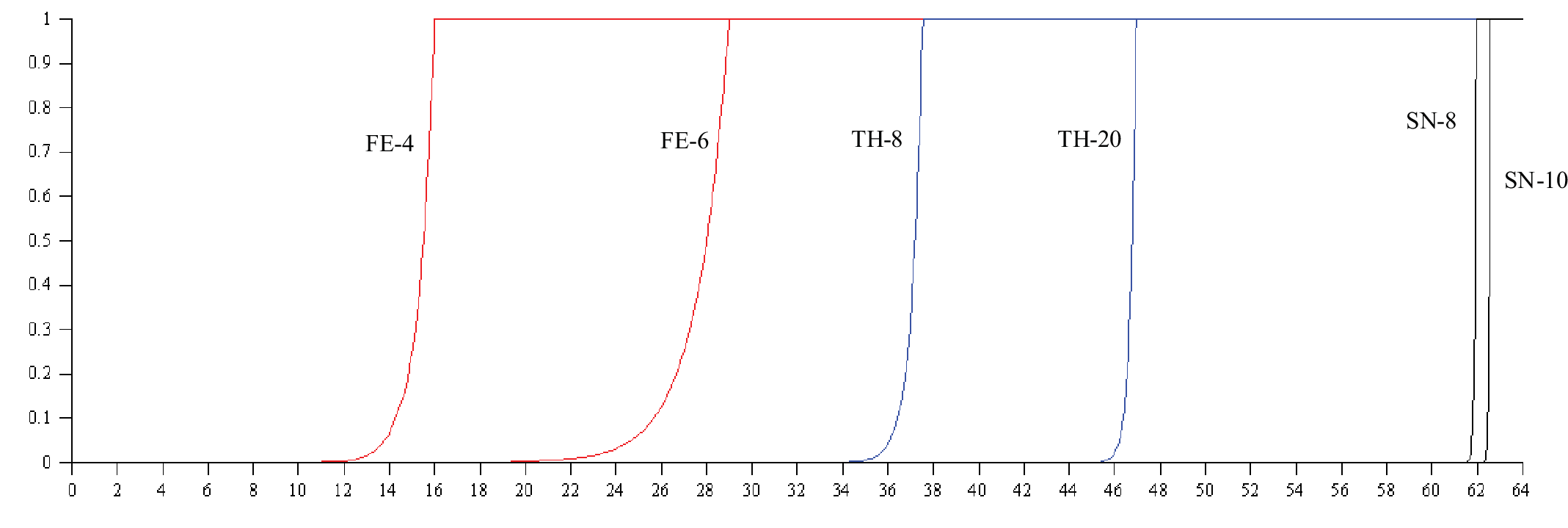}}
\else
\scalebox{0.33}{\includegraphics{graphs.eps}}
\fi
\caption{\textbf{Illustration of results.}
The message space has $N=2^{64}$ points. 
The graphs show established upper bounds on CCA advantage when the adversary 
asks~$q$ queries, where~$\log_2(q)$ labels the $x$-axis.
\textbf{Rightmost two graphs}: the new results---the \name cipher for 
either eight {passes} ($512$ rounds)
(SN-8) or 10 (SN-10), as given by \thref{T:cca_shuffle}.
(One pass is defined as $\lceil \lg N\rceil$ rounds.) 
For comparison, the \textbf{leftmost two graphs} are for balanced Feistel,
both the classical 4-round result of Luby and Rackoff \cite{lr,patarin90} (LR-4) 
and then a six-round result of Patarin (LR-6) \cite[Th.~7]{patarin11}.
The \textbf{middle two graphs} are for the Thorp shuffle, either 
with eight passes (TH-8) or 20 (TH-20), as given by \cite[Th.~5]{mrs}.
}
\label{fi-graphs}
\end{figure}

The quantitative guarantee above is far stronger than anything a balanced Feistel network can deliver. 
The only remotely comparable bound we know, retaining security to $N^{1-\epsilon}$ queries
instead of $(1-\epsilon)N$ queries, 
is the Thorp shuffle~\cite{thorp}
(or, equivalently, a maximally-unbalanced Feistel network~\cite{mrs}). 
But the known result,
establishing $\Adv^\cca_{E'}(q)\le (2q/r+1)(4nq/N)^r$ if one shuffles $N=2^n$ points for $r(4n-2)$ rounds~\cite{mrs},
vanishes by the time that $q\ge\frac{N}{4\lg N}$.
Numerically, the Thorp-shuffle bounds come out much weaker for most~$r$, $q$, and~$N$. 
See \figref{fi-graphs} for sample graphs comparing known bounds on 
balanced Feistel, the Thorp shuffle, and \name.

As a simple numerical example, \name enciphering 
64-bit strings for 1200 rounds using 
a random round function will yield 
a maximal CCA advantage of less than $10^{-10}$, even if the adversary can ask $q=2^{63}$ queries.
While the number of rounds is obviously large, 
no other construction can deliver 
a comparable guarantee, achieving security even when~$q$ is close to~$N$. 

For a more complexity-theoretic discussion of \name,
see \secref{se-complexity}.

\vspace{10pt}
\begin{wrapfigure}{l}{3.4in}{%
\vspace{-5ex}
\fbox{\parbox{3.3in}{%
\vspace{-2.0ex}
\begin{tabbing}
123\=123\=\kill
 proc $\tweak{E}_{\KF}(T, \X)$ \`\comment{\bf Tweakable \name} \\
 {for} $i\gets1$ {to} $\nr$ {do}\\
 \iffull
  \>   $\X'\gets \K_i - \X$, ~$\hatX \gets \max(\X,X')$\\
 \else
 \>   $\X'\gets \K_i - \X$\\
 \>   $\hatX \gets \max(\X,X')$\\
 \fi
 \>   {if} $\F_i(T, \hatX)=1$ then $\X \gets \X'\;$\\ 
{return} $\X$ 
\end{tabbing}
{\small \textbf{Fig.~4.} Tweakable blockcipher $\tweak{E}=\TSN[r,N, +, \TSpace]$ encrypts \mbox{$\X\!\in\!\NN$} 
under a tweak $T \in \TSpace$ using 
a key $\KF$ naming $\K_1,\ldots, \K_\nr\!\in\![N]$ and round functions 
\mbox{$\F_1,\ldots,\F_\nr\Colon \TSpace \times [N]   \rightarrow\bits$}.}
}}}
\vspace{-4ex}
\end{wrapfigure}
\heading{Adding tweaks.}
One can also turn \name to a {tweakable blockcipher}~\cite{lrw}
by adding the tweak into the scope of the round functions, 
as shown in Fig.~4. 
We generalize the CCA security to the tweakable setting as follows. 
Consider an adversary~$A$ attacking the tweakable blockcipher~$\tweak{E}$. 
The adversary is then dropped into one of two worlds. 
In the first, the adversary gets an oracle for $\tweak{E}_\KF(\cdot)$, for a random~$\KF$,
and also an oracle for its inverse, $\tweak{E}_\KF^{-1}(\cdot, \cdot)$.
Alternatively, the adversary is given a family of independent, uniformly random permutations 
$\tweak{\pi}\Colon \TSpace \times \NN  \rightarrow\NN$, along with its inverse,~$\tweak{\pi}^{-1}(\cdot, \cdot)$. 
Define 
\[\Adv^\cca_{\tweak{E}}(q) \;\;=\;\; \max_A\; 
\left\{\Pr[A^{\,\tweak{E}_\KF(\cdot, \cdot),\:\tweak{E}_\KF^{-1}(\cdot, \cdot)}\Rightarrow1] - 
  \Pr[A^{\,\tweak{\pi}(\cdot, \cdot),\:\tweak{\pi}^{-1}(\cdot, \cdot)}\Rightarrow1]
\right\},
\]
the maximum over all adversaries that ask at most~$q$ total queries.
In \secref{se-tweak} we recite follow-on work \cite{sr} that shows 
\begin{eqnarray*}
\textbf{Adv}^{\mathrm{cca}}_{\TSN[r,N,+, \TSpace]}(q) &\;\;\le\;\;& \frac{8N^{3/4}}{\sqrt{(r + 4)}} \left(\frac{q + N}{2N}\right)^{(r + 4)/8}\;.
\end{eqnarray*}
Most prior provably-secure constructions of tweakable blockciphers, such as~\cite{lrw,ro}
only achieve security up to $q = O(\sqrt{N})$ queries. 
Minematsu~\cite{min} gives the first construction of tweakable blockciphers
that goes beyond the birthday bound, 
but his method only admits inconveniently short tweaks---tweak length
must be smaller than $\lg(N) / 2$---and has to rekey the underlying blockcipher for every tweak change. 
Landecker, Shrimpton, and Terashima~\cite{lst}, 
and then Lampe and Seurin~\cite{ls} show how to generically 
lift a blockcipher~$E$ to a tweakable blockcipher~$\tweak{E}$ 
with $\textbf{Adv}^{\mathrm{cca}}_{\tweak{E}}(q) \leq \textbf{Adv}^{\mathrm{cca}}_{E}(q)
+ \sqrt{q} (q / N)^{r / 4}$, 
for any integer $r \geq 2$,  
by using about $r$ calls of~$E$. 
Their methods use $r$ fixed keys for~$E$, 
and can handle tweaks of arbitrary length. 
For applications of small domains, 
one would instantiate~$E$ from Thorp shuffling or \name
to achieve security up to only $q = O(N^{(1 - \epsilon)})$ queries, 
whereas tweakable \name works for $q = O((1 - \epsilon)N)$ queries. 
In addition, this instantiation is much slower than 
tweakable \name.

\heading{Format-preserving encryption.} 
\Name was originally invented as a solution for 
\textit{format-preserving encryption} (FPE) \cite{bs,br,brrs}, where
it provides the best known solution, in terms of 
proven-security bounds, 
when~$N$ is too big to spend linear time computing,
yet too small for conventional constructions 
to deliver desirable bounds.
This landscape has not much changed with the recent work of
Stefanov and Shi \cite{ss}, who, following Granboulan and Pornin~\cite{gp}, 
show how to speed up 
(\eg, to $\TildeTheta(N^{0.5})$ time)
determining where a card goes in a particular $N$-card shuffle 
after spending $\TildeTheta(N)$ time at key-setup.
Subsequent to our initial publication of this paper~\cite{hmr}, 
Ristenpart and Yilek~\cite{mix}
used swap-or-not as a building block to realize a particular cipher
of~$N$ cards in $\Theta(\lg^2(N))$ time. 
Their method was later refined by Morris and Rogaway~\cite{sr}
to reduce the \emph{average} running time to $\Theta(\lg(N))$
while retaining the $\Theta(\lg^2(N))$ worst-case running time. 
For more discussion of \name and its use in FPE, 
see \secref{se-fpe}.

\section{Preliminaries} \label{se-prelim}

\noskipheading{Total variation distance.} Let~$\mu$ and~$\nu$ be probability distributions on $\Omega$. 
The \emph{total variation distance} between distributions~$\mu$ and~$\nu$ is defined as
\[ \| \mu - \nu \| = \frac{1}{2} \sum_{x \in \Omega} |\mu(x)  - \nu(x)| = 
\max_{S \subset \Omega} \{ \mu(S) - \nu(S) \} \enspace.\]

\headingg{Blockciphers.}
Let $E\colon \calK \times \calM \rightarrow \calM$ be a \blockcipher, meaning that $\calK$ and~$\calM$ are finite
and each $E_K(\cdot)=E(K,\cdot)$ is a permutation on~$\calM$. 
We emphasize that~$\calK$ and~$\calM$ need not consist of binary strings of some particular length,
as is often assumed to be the case.
For any \blockcipher~$E$, we let~$E^{-1}$ be its inverse \blockcipher.

For \blockcipher 
$E \colon \calK \times \calM \rightarrow \calM$ and 
adversary~$A$ the \textit{advantage} of~$A$ in carrying out an (adaptive) 
{chosen-ciphertext attack} (CCA) 
on~$E$ is %
\[\Adv_E^\cca(A)\!=\!\allowbreak \Pr[K \getsr \calK\!\colon\!  A^{E_K(\cdot), E^{-1}_K(\cdot)} \Rightarrow 1]
\!-\! \allowbreak \allowbreak \Pr[ \pi \getsr \Perm(\calM)\!\colon\! A^{\pi(\cdot), \pi^{-1}(\cdot)} \Rightarrow 1] .\]
Here $\Perm(\calM)$ is the set of all permutations on~$\calM$. 
We say that~$A$ carries out an (adaptive) {chosen-plaintext attack} (CPA) 
if it asks no queries to its second oracle.
Adversary~$A$ is \emph{non-adaptive} if it asks the same queries on every run.  
Let $\AD{E}(q)$  be the maximum advantage of any (adaptive) CCA  
adversary against~$E$ subject to the adversary asking at most~$q$ total oracle queries.
Similarly define $\NA{E}(q)$ for nonadaptive CPA attacks (NCPA).

For \blockciphers 
$F,G\colon \calK\times \calM\rightarrow\calM$ 
let 
$F\circ G$ denote their cascade, with
$F$'s output fed into~$G$'s input; formally, 
$F\circ G\colon \calK^2\times\calM\rightarrow\calM$ 
is defined by $(F\circ G)_{(K,K')} = G_{K'}(F_K(X))$. 

\heading{Lifting ncpa to cca security.} 
We bound the CCA-security of a Feistel network from its NCPA-security 
by using the following result of Maurer, Pietrzak, and Renner~\cite[Corollary~5]{mpr}.
It is key to our approach, effectively letting us assume that our adversaries are 
of the simple, NCPA breed. 
Recall that in writing $F\circ G$, the \blockciphers are, in effect, independently keyed.

\begin{lemmaa}[Maurer-Pietrzak-Renner] %
If $F$ and $G$ are \blockciphers on the same message space 
then, for any~$q$, 
$\AD{F \circ\, G^{-1}}(q) \leq \NA{F}(q) + \NA{G}(q)$.
\label{L:composition_security}
\label{mpr}
\end{lemmaa}%

\heading{Tweakable blockciphers.}
Let $\tweak{E}\colon \calK \times \calM \times \TSpace \rightarrow \calM$ be a 
tweakable blockcipher, meaning that $\calK,\calM,$ and $\TSpace$ are finite
and each $E_K(\cdot, T)=E(K, \cdot, T)$ is a permutation on~$\calM$, for every $T \in \TSpace$. 
Thus, when~$\TSpace$ is a singleton, $\tweak{E}$ degenerates to a blockcipher.
For any tweakable \blockcipher~$\tweak{E}$, we let~$\tweak{E}^{-1}$ be its inverse 
tweakable \blockcipher.

For tweakable \blockcipher 
$\tweak{E} \colon \calK \times \calM \times \TSpace \rightarrow \calM$ and 
adversary~$A$ the \textit{advantage} of~$A$ in carrying out an (adaptive) 
{chosen-ciphertext attack} (CCA) 
on~$E$ is %
\[\Adv_{\tweak{E}}^\cca(A)\!=\!\allowbreak \Pr[K \getsr \calK\!\colon\!  A^{\tweak{E}_K(\cdot, \cdot), 
\tweak{E}^{-1}_K(\cdot, \cdot)} \Rightarrow 1]
\!-\! \allowbreak \allowbreak \Pr[ \tweak{\pi} \getsr \Perm(\calM, \TSpace)\!\colon\! 
A^{\tweak{\pi}(\cdot, \cdot), \tweak{\pi}^{-1}(\cdot, \cdot)} \Rightarrow 1] .\]
Here $\Perm(\calM, \TSpace)$ is the set of all $\TSpace$-indexed families of permutations on~$\calM$.
We say that~$A$ carries out an (adaptive) {chosen-plaintext attack} (CPA) 
if it asks no queries to its second oracle.
Adversary~$A$ is \emph{non-adaptive} if it asks the same queries on every run.  
Let $\AD{\tweak{E}}(q)$  be the maximum advantage of any (adaptive) CCA  
adversary against~$\tweak{E}$ subject to the adversary asking at most~$q$ total oracle queries.
Similarly define $\NA{\tweak{E}}(q)$ for nonadaptive CPA attacks (NCPA).

To lift NCPA security to CCA security in the tweakable setting, 
we use the following result of Lampe and Seurin~\cite{ls}. 
For tweakable blockciphers $\tweak{F}$ and $\tweak{G}$ 
of the same message space and same tweak space, 
define $\tweak{F} \circ \tweak{G}: K^2 \times \calM \times \TSpace$ by
$\tweak{F} \circ \tweak{G}_{K, K'}(X, T)
= \tweak{G}_{K'}( \tweak{F}_K(X, T), T)$. 

\begin{lemmaa}[Lampe-Seurin] %
If $\tweak{F}$ and $\tweak{G}$ are tweakable blockciphers on the same message space 
and tweak space
then, for any~$q$, 
$\AD{\tweak{F} \circ\, \tweak{G}^{-1}}(q) \leq 2\sqrt{\NA{\tweak{F}}(q)} + 2\sqrt{\NA{\tweak{G}}(q)}$.
\label{L:ls}
\end{lemmaa}%

\section{Security of \NAME} 

Fix a finite abelian group $G=([N], +)$ where $[N]=\{0,1,\ldots,N-1\}$. 
We define the \name shuffle
$\Shuffle[r,N,+]$ of~$r$ rounds over the elements of~$G$. 
The shuffling at round~$t$ is as follows. 
Initially, each of~$N$ distinct cards is at 
a position in the set~$[N]$.
To shuffle during this round,
choose $K_t \getsr [N]$, the \emph{subkey} at round~$t$. 
Then, for each set $\{X, K_t - X\}$ with $X\in G$, choose $b \getsr \bits$
and then swap the cards at positions~$X$ and $K_t - X$ if $b = 1$. 

Let $\{\deck_t : t \geq 0\}$ be the Markov chain representing
the \name shuffle with $N$ cards. More 
formally, 
let $\cards$ be a set of 
cardinality $N$, 
whose elements we call \emph{cards}.
The state space of $\{\deck_t\}$ is the set of bijections
from $\cards$ to $\{0,\dots,N-1\}$. 
For a card $\card \in \cards,$ 
we interpret $\deck_t(\card)$ as the position of card $\card$
at time $t$. 

Let $A$ be a deterministic adversary that makes exactly $q$~queries.
Our proof is based on an analysis of the mixing rate of the \name 
shuffle. However,
since~$A$
makes only $q \le N$~queries, we need only bound the rate at which 
some $q$-element subset of the cards mixes. 
So let $\card_1, \dots, \card_q$ be distinct cards in~$\cards$, and let 
$\mT_t$ be 
the vector of
positions of cards $\card_1, \dots, \card_q$ at time $t$. 
For~$j$ in $\{1,\dots,q\}$ we write
$\mT_t(j)$ for the position of card $\card_j$ at time $t$, and define
$\mT_t(1, \dots, j) = 
(\mT_t(1), \dots, \mT_t(j))$.
We shall call~$\mT_\n$ the {\it projected \name shuffle.}
Note that the stationary distribution of $\mT_\n$,
which we denote by $\mpi$, is uniform 
over the set of distinct $q$-tuples of elements from~$G$. 
Equivalently, $\mpi$
is the distribution
of $q$ samples without replacement from $G$.
Let $\mtau_\n$ denote the distribution of $\mT_\n$.

\begin{theorem}[Rapid mixing]  \label{thm:thorp-nCPA-markov} \label{th-main-markov}
Consider the \name shuffle $\SN[r,N,+]$ for $r,N\ge1$, and let $q \in \{1,\dots,N\}$. 
Fix $z_1,\ldots,z_q$ and
let \mbox{$\{\mT_t\colon t \ge 0\}$} be the
  corresponding projected \name shuffle, 
  let~$\mpi$ be its stationary distribution, and let~$\mtau_{t}$ be the 
  distribution of $\mT_t$.
  Then
\begin{equation*}
  \Vert \mtau_{r} - \mpi \Vert \le 
  \frac{2N^{3/2}}{r+2} \left( \frac{ q + N     }{2N}\right)^{r/2 + 1}\;.
\end{equation*}

\end{theorem}

\begin{proof}
Let $\mtau_t^k$ be the conditional distribution of $\mT_t$ given the 
subkeys $K_1, \dots, K_{r}$. 
(Here we consider $K_1, \dots, K_r$ random variables, 
and we condition on the $\sigma$-algebra of these random variables.) %
We will actually show that $\Exp(\Vert \mtau_r^k - \mpi \Vert)$ satisfies the claimed inequality. 
Note that since $K_1,\ldots,K_r$ are random
variables, so is $\tau_r^k$, and hence so is 
$\Vert \tau_r^k-\pi\Vert$.
This implies the theorem since 
$\mtau_r = \Exp(\mtau_r^k)$ and hence
\begin{eqnarray*}
\Vert \mtau_r - \mpi \Vert 
   =    \Vert \Exp (\mtau_r^k - \mpi) \Vert 
   \leq \Exp \Bigl( \Vert \mtau_r^k - \mpi \Vert \Bigr),
\end{eqnarray*}
by Jensen's inequality, since for distributions $\mu$ and $\tau$, the 
total variation distance $\Vert \mu - \tau \Vert$ is half the $L^1$-norm 
of $\mu - \tau$, and the $L^1$-norm is convex. 
For a distribution~$\nu$ on $q$-tuples of $\Omega$, define
\begin{eqnarray*}
  \nu(u_1,\dots, u_{j}) &=& \Pr[Z_1=u_1, \dots, Z_j=u_j] \mbox{ and}\\
  \nu(u_j \given u_1,\dots, u_{j-1}) &=& \Pr[Z_j=u_j \given Z_1=u_1, \dots, Z_{j-1}=u_{j-1}]
\end{eqnarray*}
where $(Z_1, \dots, Z_q) \sim \nu$. 
For example, $\mtau_\n(u_1, \dots, u_j)$ is the probability
that, in the \name shuffle,
cards $\card_1, \dots, \card_j$ land in positions $u_1, \dots, u_j$ at time~$t$, while
$\mtau_\n(u_j \given u_1,\dots, u_{j-1})$ is the probability that at time~$t$
card~$\card_j$ is in position~$u_j$ given that 
cards $\card_1, \dots, \card_{j-1}$ are in positions $u_1, \dots, u_{j-1}$.
On the other hand, $\mpi(u_j \given u_1,\dots u_{j-1})$ is the probability that,
in a uniform random ordering,
card~$\card_j$ is in position~$u_j$ given that cards $\card_1, \dots, \card_{j-1}$
land in positions $u_1, \dots, u_{j-1}$.

Each of the 
conditional distributions
$\mtau_\n^k(\;\cdot \given u_1,\dots, u_{j-1})$ 
converges to uniform as $\n \to \infty$. When all of 
these distributions are ``close'' to uniform, then $\mtau_\n^k$
will be close to $\mpi$. In fact, we only need
the conditional distributions to be close ``on average,'' as
is formalized in 
the following lemma, which is easily established using coupling.
For a proof, see \cite[Appendix~A]{mrs}.  %

\begin{lemmaa}
\label{tv} \label{le-main}
Fix a finite nonempty set $\Omega$ and 
let~$\mu$ and~$\nu$ be probability distributions supported on 
$q$-tuples of elements of $\Omega$, and  
suppose that $(Z_1, \dots, Z_{q}) \sim \mu$.
Then
\begin{equation}
\label{fe}
\Vert \mu - \nu \Vert \;\leq\; \sum_{l=0}^{q-1} 
\Exp  \Bigl( \Vert \mu(\; \cdot \given Z_1, \dots, Z_\ell) - 
\nu(\; \cdot \given Z_1, \dots, Z_\ell) \Vert \Bigr)\,.
\end{equation}
\end{lemmaa}

\noindent
Note that in the above lemma, since $Z_1, \dots, Z_q$ are random variables 
(whose joint distribution is given by $\mu$), 
so is
$\Vert \mu( \;\cdot\given Z_1, \dots, Z_\ell     ) - 
\nu( \;\cdot\given Z_1, \dots, Z_\ell) \Vert$ for 
every $\ell < q$; each summand in the right-hand side
of (\ref{fe}) is the expectation of one of these random variables.

Recall that $\mtau_t^k$ is the conditional distribution of $\mT_t$
given $K_1, \dots, K_r$.
Fix $\ell \in \{0, \ldots, q - 1\}$. 
We wish to 
bound the expected distance between 
the distribution
$\mtau_t^k(\; \cdot \given \mT_t(1), \dots, \mT_t(\ell))$ and 
$\mpi(\; \cdot \given \mT_t(1), \dots, \mT_t(\ell))$
(\ie, the uniform distribution on $G \setminus \{\mT_t(1), \dots, \mT_t(\ell)\}$).

For $t \geq 0$, let $S_t = G \setminus \{\mT_t(1), \dots, \mT_t(\ell)\}$.
Thus $S_t$ is the set of positions
that card $z_{\ell+1}$ could be located in at time~$t$, given the positions
of cards $z_1,...,z_\ell$.
For $a \in S_t$, 
let $p_t(a) = \mtau_t^k(a \given \mT_t(1), \dots, \mT_t(\ell))$.
Then we have
\begin{equation}
\label{eq:distance}
\Vert 
\mtau_t^k(\; \cdot \given \mT_t(1, \dots,\ell)) - 
\mpi(\; \cdot \given \mT_t(1, \dots,\ell)) \Vert 
= \frac{1}{2}\sum_{a \in S_t} |p_t(a) - 1/\LL|,
\end{equation}
where $\LL = |S_t| = N - \ell$. 
Using the Cauchy-Schwarz inequality twice gives 
\begin{eqnarray}
 \left( \Exp\left[\;\sum_{a \in S_t} |p_t(a) - 1/\LL| \right] \right)^2 
&\leq& \Exp\left[\;  \left(\sum_{a \in S_t} |p_t(a) - 1/\LL| \right)^2\right] \nonumber\\
&\leq& \LL \cdot \Exp\left[\;\sum_{a \in S_t} (p_t(a) - 1/\LL)^2 \right] \nonumber \\
&\leq& N \cdot \Exp\left[\;\sum_{a \in S_t} (p_t(a) - 1/\LL)^2 \right] \enspace. \label{eq:cauchy}
\end{eqnarray}
We shall prove, by induction on~$t$, that 
\begin{equation}
\Exp\left[\; \sum_{a \in S_t} (p_t(a) - 1/\LL)^2 \right] \leq \left(\frac{\ell + N}{2N}\right)^{t}
\label{eq:induction}
\end{equation}
for every $t \leq r$. Then, 
substituting $t = r$ to \eqref{eq:distance}, \eqref{eq:cauchy}, 
and \eqref{eq:induction}, 
we have
\begin{eqnarray}
& &\Exp \Bigl(\Vert 
\mtau_r^k(\; \cdot \given \mT_r(1, \dots,\ell)) - 
\mpi(\; \cdot \given \mT_r(1, \dots,\ell)) \Vert \Bigr) \nonumber \\ 
&\leq&
\frac{1}{2}\left(N \cdot \Exp\left[\;\sum_{a \in S_r} (p_r(a) - 1/\LL)^2 \right] \right)^{1/2}
\leq
\frac{\sqrt{N}}{2} \left(\frac{\ell + N}{2N}\right)^{r/2} \enspace. \label{eq:bc}
\end{eqnarray}
Substituting this into \lemref{tv} gives
\begin{eqnarray*}
\Exp \Bigl( \Vert \mtau_r^k - \mpi \Vert \Bigr) &\leq&
\sum_{\ell = 0}^{q-1} 
\Exp \Bigl(\Vert 
\mtau_r^k(\; \cdot \given \mT_r(1, \dots,\ell)) - 
\mpi(\; \cdot \given \mT_r(1, \dots,\ell)) \Vert \Bigr) \\
&\leq& \sum_{\ell = 0}^{q - 1}
\frac{\sqrt{N}}{2} \left(\frac{\ell + N}{2N}\right)^{r/2} 
\\
&\leq& N^{3/2} \int_0^{q/2N} (1/2 + x)^{r/2} dx
\leq
\frac{2N^{3/2}}{r + 2} \left(\frac{q + N}{2N}\right)^{r/2 + 1} \enspace.
\end{eqnarray*}

We now verify equation \eqref{eq:induction}. 
First, consider the base case $t = 0$. 
Since the initial positions of the cards are deterministic, 
\[ \Exp\Bigl[ \sum_{a \in S_0} (p_0(a) - 1/\LL)^2 \Bigr]
= (1 - 1/\LL)^2 + (\LL - 1) \cdot (0 - 1/\LL)^2
= 1 - 1/\LL < 1\enspace.\]
Now suppose that equation \eqref{eq:induction} holds for~$t$. 
We prove that it also holds for $t + 1$. 
Define
$s_t = \sum_{a \in S_t} \left(p_t(a) - 1/\LL\right)^2$.  
It is sufficient to show that
\begin{equation}
\label{ver}
\Exp(s_{t+1} \given s_t) = \left( {\ell + N \over 2N} \right) s_t.
\end{equation}
Define $f: S_t \to S_{t+1}$ by 
\[
f(a) =
\left\{\begin{array}{ll}
a &
\mbox{if $a \in S_{t+1}$;} \\
K_{t + 1} - a \;\;\;\; & \mbox{otherwise.}\\
\end{array}
\right.
\]
Note that $f$ is a bijection from $S_t$ to $S_{t+1}$: it sends $S_t$ to $S_{t+1}$ because
if $a \in S_t$ then either $a$ or $K_{t+1} - a$ must be in  $S_{t+1}$, and it has 
an inverse $f^{-1}: S_{t+1} \to S_t$ defined by 
\[
f^{-1}(b) =
\left\{\begin{array}{ll}
b &
\mbox{if $b \in S_{t}$;} \\
K_{t + 1} - b \;\;\;\; & \mbox{otherwise.}\\
\end{array}
\right.
\]
Furthermore, note that
\[
p_{t+1}(f(a)) =
\left\{\begin{array}{ll}
p_t(a) &
\mbox{if $K_{t + 1} - a \notin S_t$;} \\[4pt]
\half p_t(a) + \half p_t(K_{t + 1} - a) \;\;\;\;& \mbox{otherwise.}\\
\end{array}
\right.
\]
Since $K_{t+1}$ is independent of the process up to time $t$, 
\iffull
we have $\Pr[ K_{t + 1} - a = y \given s_t] = 1/N$ for every $y \in G$. 
\else
for every $y \in G$, we have $\Pr[ K_{t + 1} - a = y \given s_t] = 1/N$.
\fi
Hence, since $|S_t| = \LL$, conditioning on the value of $K_{t + 1} - a$ gives 
\begin{equation}
\label{varbound}
 \Exp\Bigl( \Bigl[p_{t+1}(f(a)) - \sfrac{1}{\LL}\Bigr]^2 \given s_t\Bigr)
={\ell \over N} \left(p_t(a) - \hspace{-0.5ex}\sfrac{1}{\LL}\right)^2
+ {1\over N} \sum_{y \in S_t} \Bigl[ {p_t(a) + p_t(y) \over 2} - \sfrac{1}{\LL} \Bigr]^2.
\end{equation}
The sum can be rewritten as
\begin{align*}
&  \sum_{y \in S_t} \quarter \Bigl[ \left(p_t(y) - 1 / \LL\right) + \left(p_t(a) - 1 / \LL\right) \Bigr]^2\\
=  
&\quarter \hspace{-0.5ex}\sum_{y \in S_t} \hspace{-0.25ex}
   \left(p_t(y) - \hspace{-0.5ex} 1/ \LL\right)^2 + \hspace{-0.15ex}
\half  
\left(p_t(a) - \hspace{-0.5ex} 1 / \LL\right) \hspace{-0.25ex}\sum_{y \in S_t} \hspace{-0.25ex}\left(p_t(y) - \hspace{-0.5ex} 1 / \LL\right) 
+\hspace{-0.5ex}
\quarter \hspace{-0.5ex}
\sum_{y \in S_t} \hspace{-0.25ex}
(p_t(a) - 1/\LL)^2   \\
= &\quarter s_t + \sfrac{\LL}{4} \left(p_t(a) - 1/\LL\right)^2,
\end{align*}
since $\sum_{y \in S_t} \left(p_t(y) - 1/ \LL\right) = 0$.  
Combining this with (\ref{varbound}) 
gives
\begin{equation}
\label{varboundt}
\Exp\Bigl( \Bigl[p_{t+1}(f(a)) - 1 / \LL \Bigr]^2 \given s_t\Bigr) = {s_t \over 4N}
+ {4\ell + \LL \over 4N} \left(p_t(a) - 1 / \LL\right)^2.
\end{equation}
Note that
\begin{eqnarray*}
\Exp(s_{t+1} \given s_t) &=& 
\sum_{b \in S_{t+1}} \Exp\Bigl( \Bigl[p_{t+1}(b) - 1/ \LL\Bigr]^2  \given s_t \Bigr) \\
&=& 
\sum_{a \in S_{t}} \Exp \Bigl(
\Bigl[p_{t+1}(f(a)) - 1/\LL \Bigr]^2 \given s_t \Bigr) .
\end{eqnarray*}
Evaluating each term in the sum using (\ref{varboundt}) 
gives
\begin{eqnarray*}
\Exp(s_{t+1} \given s_t) &=& 
{\LL s_t \over 4N} + 
{4\ell + \LL \over 4N} \sum_{a \in S_t} \left(p_t(a) - 1 / \LL \right)^2 \\
&=& {\LL s_t \over 4N} + {(4\ell + \LL) s_t \over 4N} \\
&=& {\ell + N \over 2N} s_t,
\end{eqnarray*}
where the last line holds because $\LL + \ell = N$. 
It follows that $\Exp(s_{t+1} \given s_t) = \Bigl({\ell + N \over 2N}\Bigr) s_t$,
which verifies~(\ref{ver}) and hence~(\ref{eq:induction}). 
This
completes the proof.
\qed
\end{proof}

\heading{CCA-security.}
Observe that if $E = \Shuffle[r, N, +]$
for some abelian group $G=([N], +)$ then $E^{-1}$ is also $\Shuffle[r, N, +]$.
Employing~\lemref{mpr} we conclude our main theorem.

\begin{theorem} \sl 
Let $E = \Shuffle[2r, N, +]$. 
Then 
$\displaystyle\AD{E}(q) \leq \frac{4N^{3/2}}{r + 2} \left(\frac{q + N}{2N}\right)^{r/2 + 1}$. 
\label{T:cca_shuffle}
\end{theorem}

\section{Complexity-Theoretic Interpretation} 
\label{se-conclusions}
\label{se-complexity}

While \thref{T:cca_shuffle} is information-theoretic,
it should be clear that the result applies to the complexity-theoretic setting too, 
in exactly the same manner as Luby-Rackoff~\cite{lr} and its successors.
Namely, from a PRF $F\Colon\calK\times\bits^*\rightarrow\bits$ and a number~$n$, 
define $n$-bit round functions $F_i(X)$ whose $j$th bit is $F(\code{i,j,n,X})$. 
Also define $n$-bit round keys~$K_i$ whose $j$th bit is $F(\code{i,j,n})$.
Using these components,
apply the \name construction for, say, $r=7n$ rounds, yielding a PRP~$E$ on~$n$ bits.
Translating the information-theoretic result into this setting,
the PRP-security of~$E$ is the PRF-security of~$F$ 
minus a term that remains negligible until
$q=(1-\epsilon)2^n$ adversarial queries, for any $\epsilon>0$. 
That is, from the asymptotic point of view,
the \name construction preserves essentially all 
of a PRF's security in the constructed~PRP. 

We emphasize that our security results only cover the (strong) PRP notion of security. 
An interesting question we leave open is whether the \name cipher 
is indifferentiable from a random permutation~\cite{mrh}.  
Following Coron, Patarin, and Seurin~\cite{cps}, 
Holenstein, K\"{u}nzler, and Tessaro show that the 14-round Feistel construction 
is indifferentiable from a random permutation~\cite{hkt}. 
But their proof is complex and delivers very poor concrete-security bounds.
It would be desirable to have a construction supporting a simpler proof with better bounds.

\section{Incorporating Tweaks}\label{se-tweak}

\noskipheading{NCPA security.}
Subsequent to our initial publication of the current work \cite{hmr}, 
Morris and Rogaway~\cite{sr} showed 
that tweakable swap-or-not is a good tweakable blockcipher. 

\begin{theorem}[\cite{sr}] \sl 
Let $\tweak{E} = \TSN[r, N, +, \TSpace]$. 
Then 
$\displaystyle\NA{\tweak{E}}(q) \leq \frac{2N^{3/2}}{r + 2} \left(\frac{q + N}{2N}\right)^{r/2 + 1}$. 
\end{theorem}

\heading{Handling long tweaks efficiently.} 
The construction  in Fig.~4 will be slow for very long tweaks. 
We can improve the speed by using a PRF
$H\Colon \calK \times \bits^* \to \bits^{128}$. 
A key $\KF$ names $K_1, \ldots, K_r \in [N]$, round functions $F_1, \ldots, F_r\Colon [N] \times [M] \to \bits$, 
and a key $K \in\calK$. 
At each round, instead of computing $F_i(T, \hat{X})$,  
we'll use $F_i(H_K(T), \hat{X})$. 
For small and moderate domains $(N \leq 2^{128})$, 
this construction will use at most two AES calls per round---the preprocessing cost of $H_K(T)$
is discounted.

\heading{CCA security.}
Observe that if $\tweak{E} = \TSN[r, N, +, \TSpace]$
for some abelian group $G=([N], +)$ then $\tweak{E}^{-1}$ is also $\TSN[r, N, +, \TSpace]$.
Employing~\lemref{L:ls} we conclude the following.

\begin{theorem} \sl 
Let $\tweak{E} = \TSN[2r, N, +, \TSpace]$. 
Then 
$\displaystyle\AD{\tweak{E}}(q) \leq \frac{8N^{3/4}}{\sqrt{2r + 4}} \left(\frac{q + N}{2N}\right)^{(r + 2)/4}$. 
\end{theorem}

\section{Format-Preserving Encryption} \label{se-fpe}

In the \textit{format-preserving encryption} (FPE) problem,
one wants to encipher on an arbitrary set~$\calX$, often $\calX=[N]$ for some number~$N$.
Usually constructions are sought that start from a conventional blockcipher, like AES.
The problem has attracted increasing interest 
\cite{br,brrs,mrs,su,bs,nbs,video,gp,ss,bps,brs}, 
and is the subject of ongoing standardization work by NIST and the IEEE.

When~$N$ is sufficiently small that one can afford $\Tilde\Omega(N)$-time
to encrypt, provably good solutions are easy, by directly realizing a random shuffle~\cite{br}.
And when~$N$ is sufficiently large that no adversary could ask anything near $N^{1/2}$ queries,
nice solutions are again easy, using standard cryptographic constructions
like multi-round Feistel.
But for intermediate-size domains, 
like those with \mbox{$2^{30}$--$2^{60}$} points, the 
bounds associated to well-known construction are disappointing, even if 
known attacks are not remotely feasible, and
spending time proportional to the 
domain size, even in key-setup phase, is not attractive. 

With these problematic-size domains in mind, 
suppose we use \name to 
encipher \mbox{9-digit} social security numbers $(N \approx 2^{30})$.
Employing \thref{T:cca_shuffle},
if we use 340 rounds %
we are guaranteed 
a maximal CCA advantage of less than $10^{-10}$ 
even if the adversary can ask $q=10^{8}$ queries.
Similarly, suppose we use \name to 
encipher \mbox{16-digit} credit cards ($N\approx\!2^{53}$).
If we use 500 rounds %
we are guaranteed 
a maximal CCA advantage of less than $10^{-10}$
even if the adversary can ask $q=10^{15}$ queries.
(Of course these numbers assume random round functions; if one bases the 
construction on AES, say, one will have to add in a term for its
insecurity.)
The round counts are obviously high, yet the rounds are fast and the
guarantees are strong. 
(We note too that, at least for the binary-string setting and AES as a 
starting point, there are
tricks to reduce the number of \blockcipher calls by a factor of five, 
as shown in prior work \cite{mrs}. But this is probably not 
helpful in the presence of good AES support, 
as with recent Intel processors.) 

A very different approach to small-domain FPE 
is taken by 
Granboulan and Pornin \cite{gp}, who show how to 
realize a particular shuffle on~$N$ cards 
in $O(\lg^3 N)$ encryption time and $O(\lg N)$ space.
But the method seems to be impractical, requiring 
extended-precision arithmetic to sample from 
a hypergeometric distribution.
Stefanov and Shi go on to show how to
exploit preprocessing to realize a different $N$-card 
shuffle \cite{ss}.
Their method is applicable when 
the key-setup cost of $\TildeTheta(N)$ is feasible, 
as is key storage and per-message encryption cost of
$\TildeTheta(N^{1/2})$.
Near or beyond $N\approx 2^{30}$, 
these assumptions seem unlikely to hold in most settings.
That said, the approach 
allows an adversary to query all~$N$ points,
whereas the shuffle of this paper has only 
been proven to withstand $(1-\epsilon)N$ queries.
(We conjecture that \name works well for~$N$ 
queries and reasonable~$r$---that its mixing time is fast---but no such
result is proven here.)

Subsequent to our work initial work \cite{hmr}, Ristenpart and Yilek~\cite{mix}
used swap-or-not to replace the expensive hypergeometric samplings
in Granboulan and Pornin's construction~\cite{gp}. Not only does this cheapen the design
but also reduces the number of rounds from $\Theta(\lg^3(N))$ to $\Theta(\lg^2(N))$. 
Morris and Rogaway~\cite{sr} refined this method so that the number of round, on the average case, 
is only $\Theta(\lg(N))$, but the worst-case remains~$\Theta(\lg^2(N))$.

\section{Confusion/Diffusion Ciphers} 

\Name can also be construed as an 
approach for making a confusion/diffusion \blockcipher.
In doing this one would instantiate 
round functions
\mbox{$\F_i\!:\bits^n\rightarrow\bits$}
by a fast, concrete construction.
Choosing a good instantiation for $F_i$ is tricky. 
In the proceedings version of this paper \cite{hmr} we suggested that
a simple, plausible instantiation was to have~$\F_i$ be specified by
an $n$-bit string~$L_i$ and let  %
$\F_i(\hatX) = L_i\ip\,\hatX$
be the inner-product of~$L_i$ and~$\hatX$.
Almost immediately, both Henri Gilbert and Serge Vaudenay pointed out that 
this function being linear would result in the entire construction being trivially breakable~\cite{g,v}. 
We leave it as a problem for cryptanalysts to investigate 
what (nonlinear) choices of~$F_i$ might work well, 
and how large~$\nr$ needs to be for them. 
\subsection*{Acknowledgments}
The authors gratefully acknowledge comments from 
Mihir Bellare, Henri Gilbert,  
Terence Spies, and Serge Vaudenay.
This work was supported under 
NSF grants
DMS-1007739 and 
CNS-0904380.

\end{document}